\documentclass[%
reprint,
superscriptaddress,
amsmath,
amssymb,
aps,
prl,
longbibliography
]{revtex4-2}

\usepackage{amsmath}
\usepackage{amsfonts}
\usepackage{amssymb}
\usepackage{array}
\usepackage{bbm}
\usepackage{bm}
\usepackage{bbold}
\usepackage{braket}
\usepackage{bigints}
\usepackage{dsfont}
\usepackage[dvipsnames]{xcolor}
\usepackage{epsfig}
\usepackage{epstopdf}
\usepackage{pdfpages}
\usepackage{pgffor}
\usepackage{graphicx}
\usepackage{multirow}
\usepackage{mathtools}
\usepackage[T1]{fontenc}
\usepackage{times} 
\usepackage{tensor}
\usepackage[utf8]{inputenc}
\usepackage[unicode=true,breaklinks=true,colorlinks=true]{hyperref}

\hypersetup{citecolor=blue,urlcolor=blue,linkcolor=black}

\makeatletter 
\AtBeginDocument{\let\LS@rot\@undefined} 
\makeatother

\usepackage[normalem]{ulem}

\begin{document}

\title{Strong quantum metrological limit from many-body physics}

\author{Yaoming Chu}
\thanks{These authors contributed equally.}
\author{Xiangbei Li}
\thanks{These authors contributed equally.}
\affiliation{School of Physics, International Joint Laboratory on Quantum Sensing and Quantum Metrology, Hubei Key Laboratory of Gravitation and Quantum Physics, Institute for Quantum Science and Engineering, Wuhan National High Magnetic Field Center, Huazhong University of Science and Technology, Wuhan 430074, China}
\author{Jianming Cai}
\email{jianmingcai@hust.edu.cn}
\affiliation{School of Physics, International Joint Laboratory on Quantum Sensing and Quantum Metrology, Hubei Key Laboratory of Gravitation and Quantum Physics, Institute for Quantum Science and Engineering, Wuhan National High Magnetic Field Center, Huazhong University of Science and Technology, Wuhan 430074, China}
\affiliation{Shanghai Key Laboratory of Magnetic Resonance, East China Normal University, Shanghai 200062, China}
\begin{abstract}
Surpassing the standard quantum limit and even reaching the Heisenberg limit using quantum entanglement, represents the Holy Grail of quantum metrology. However, quantum entanglement is a valuable resource that does not come without a price. The exceptional time overhead for the preparation of large-scale entangled states raises disconcerting concerns about whether the Heisenberg limit is fundamentally achievable. Here we find a universal speed limit set by the Lieb-Robinson light cone for the quantum Fisher information growth to characterize the metrological potential of quantum resource states during their preparation. Our main result establishes a strong precision limit of quantum metrology accounting for the complexity of many-body quantum resource state preparation and reveals a fundamental constraint for reaching the Heisenberg limit in a generic many-body lattice system with bounded one-site energy. It enables us to identify the essential features of quantum many-body systems that are crucial for achieving the quantum advantage of quantum metrology, and brings an interesting connection between many-body quantum dynamics and quantum metrology.
\end{abstract}

\maketitle

Quantum metrology, as part of the rapidly rising field of quantum science and technology, is capable of achieving enhanced precision measurement by exploiting quantum strategies and promises unprecedented applications in  basic science and technology \cite{Giovannetti2011,Paris2011,Degen2017,Budker2007,Aasi2013,Backes2021}. Importantly, irrespective of classical accidental errors \cite{Giovannetti2011}, quantum mechanics imposes a fundamental limit on the accuracy to which measurements can be performed, called the Heisenberg limit \cite{Giovannetti2004,Boixo2007,Gorecki2020}. It conventionally refers to a measurement precision scaling as $1/N$ (where $N$ is typically regarded as the number of probes employed) and acts as a hallmark of scaling improvement over the standard quantum limit (SQL), which arises from uncorrelated measurements and is given by $1/\sqrt{N}$. The advantage of quantum metrology in terms of reaching the Heisenberg limit often stems from quantum entanglement \cite{Toth2014,
Nagata2007,Pezze2018}, such as the “cat” state of $N$ probes. Although the improved scaling has been unfortunately demonstrated to be fragile when suffering from noise sources \cite{Huelga1997,Escher2011,Demkowicz2012}, a number of clever noise-robust quantum metrological schemes \cite{
Kessler2014,Arrad2014,Zhou2018,Yamamoto2022} have been proposed to battle against decoherence, making it not a fundamentally unsolvable obstacle for reaching the Heisenberg limit.
%


%
In contrast, the preparation of large-scale entangled state itself is usually a highly challenging task, the required time of which may be restricted by fundamental laws. Such a state preparation time and the corresponding measurement repetition number would have a non-trivial impact on the precision limit.
Subsequently, the following intricate and critical problem remains elusive: whether is there any fundamental constraint that prevents reaching the Heisenberg limit imposed by the complexity of state preparation \cite{Dooley2016,Hayes2018,McGuinness2021}? Addressing this problem and establishing a stronger precision limit would put comparisons of entanglement-enhanced quantum measurements with the SQL on a more fair footing, and consistently consolidate the foundation of the quantum advantage offered by quantum metrology over classical counterparts. It also relates to another crucial problem: is there any simple and universal principle to find quantum metrological systems that are favorable for quantum metrology?
In this Letter, we address these problems by quantifying the growth of quantum Fisher information (QFI) for the metrological state preparation in generic quantum many-body systems with bounded one-site energy. As our main result, we find a universal speed limit for the QFI growth set by the Lieb-Robinson (LR) light cone from many-body physics \cite{Lieb1972,Hastings2006}. This provides a versatile tool to lower bound the time required by metrological resource state preparation, and thus establishes the precision limit for entanglement-enhanced quantum metrology explicitly involving the complexity of state preparation (dubbed “strong precision limit”). Applying our result to quantum many-body systems with Ising and dipolar interactions, we illustrate how to achieve the advantage of quantum metrology through long-range interacting and higher-dimensional systems, which presents insightful design principles for quantum metrological systems. The result also connects quantum metrology with quantum information propagation and offers new insights into many-body quantum dynamics via the concept of QFI growth.
\begin{figure}[t]
\centering
\includegraphics[width=86mm]{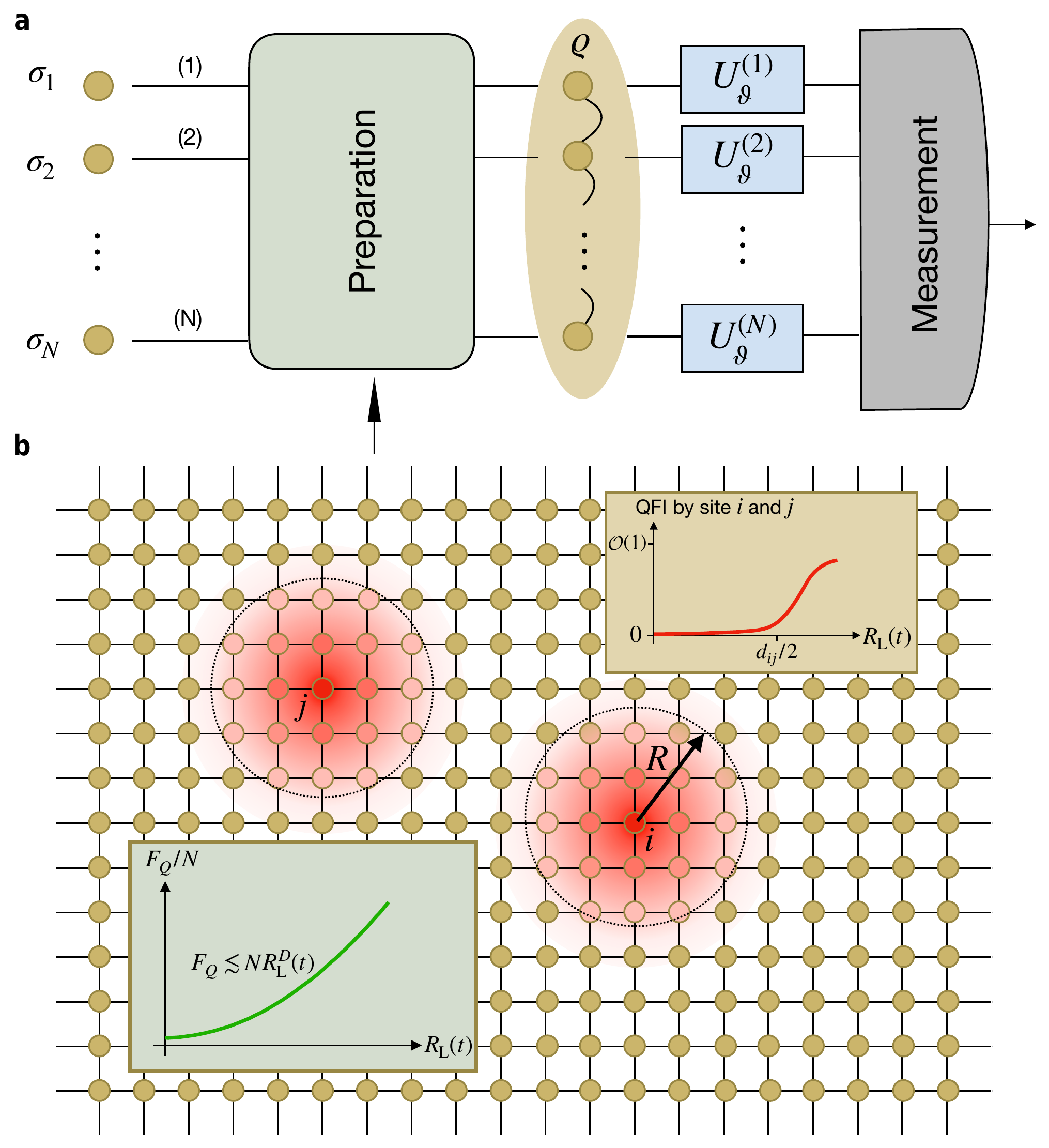}
\caption{{\bf Metrological resource state preparation and QFI growth.} {\bf a}, Conventional scheme for quantum metrology. A $N$-probe quantum system is prepared in the resource state $\varrho$ and is fed into $N$ parallel local channels (each one is described as $U_\vartheta^{(i)}\equiv e^{-i\vartheta K_i}$) to sense an unknown phase $\vartheta$. {\bf b}, Spreading of local operators in a quantum lattice. The quasilocality implied by LR bound ensures that $K_i(t)$ (red region) is well approximated by an operator $K_i(t,i[R])$ defined on a ball region $i[R]$ (dashed circle) with $R\gtrsim R_{\mathrm{L}}(t)$, where $R_{\mathrm{L}}(t)$ represents the effective light cone. The QFI collaboratively contributed by the site $i$ and $j$ (red curve) begins to be nonzero only when the two balls $K_i(t,i[R])$ and $K_j(t,j[R])$ grow to touch each other. Consequently, the upper bound of the QFI growth (green curve)  would be constrained by the effective light cone $R_{\mathrm{L}}(t)$.}
\label{Fig:model}
\end{figure}

{\it Fundamental constraint for QFI growth.---} Quantum Fisher information, $F_Q$, as a fundamental concept in quantum metrology, determines the optimal measurement precision using a given quantum resource state $\varrho$   \cite{Braunstein1994}. It quantifies the sensitivity of $\varrho$ to a parameter-dependent unitary transformation generated by an interrogation operator $\mathcal{K}$, i.e. $\varrho_\vartheta=e^{-i\vartheta \mathcal{K}} \varrho e^{i\vartheta \mathcal{K}}$ with $\vartheta$ a phase to be estimated. Usually the interrogation operator is local, $\mathcal{K}=\sum_{i=1}^N K_i$, with $K_i$ defined on the $i$-th  individual probe, see Fig.\,\ref{Fig:model}a. For a general resource state, spectrally decomposed as $\varrho=\sum _n p_n |n\rangle\langle n|$, the QFI takes a structure of the following form
\begin{equation}
\label{Eq:QFIdef}
F_{\mathrm{Q}}[\varrho,\mathcal{K}]=2 \sum_{n, m} \frac{\left(p_n-p_m\right)^2}{p_n+p_m}\left|\left\langle n|\mathcal{K}| m \right\rangle\right|^2,
\end{equation}
where the sum includes only terms with $p_n+p_m>0$. A larger QFI indicates a better distinguishability between neighboring parametrized quantum states $\varrho_\vartheta$ and $\varrho_{\vartheta+\delta\vartheta}$, and thus a more sensitive $\varrho$ to the operator $\mathcal{K}$. 

Quantum metrology often seeks such metrological states $\varrho$ (e.g. GHZ state \cite{Leibfried2004}, squeezed spin states \cite{Kitagawa1993,Bao2020}, critical ground states \cite{Frerot2018}, etc.) to enhance the optimal measurement precision. These strongly correlated quantum metrological states are usually prepared via coherent interactions between individual probes, see e.g. \cite{Zhang2005,Liu2011,Strobel2014,Song2019,Liu2022,Marciniak2022,Graham2022}. We consider generic Hamiltonians with few-body interactions on a $D$-dimensional lattice, with $\Lambda$ the total particle set and the cardinality $|\Lambda|=N$. In detail, the interaction Hamiltonian of a $k$-local form with bounded one-site energy \cite{Kuwahara2022} can be expressed as, 
\begin{equation}
\label{Eq:GeneralH}
H=\sum_{|X| \leq k} h_X, \quad \max _{i \in \Lambda} \sum_{X: X \ni i}\left\|h_X\right\| \leq \mathcal{O}(1),
\end{equation}
where each interaction term $h_X$ acts on the particle subset $X\subset \Lambda$, and $\|\bullet\|$  is the operator norm. Remarkably, in state-of-the-art quantum platforms including trapped-ion crystals \cite{Britton2012,Bohnet2016}, Bose-Einstein condensates \cite{Cirac1998,Milburn1997,Gross2010,Riedel2010,Muessel2014,Luo2017,Li2022,Muniz2020}, cavity-QED systems \cite{Hosten2016,Colombo2022,Greve2022,Leroux2010}, Rydberg-atom arrays \cite{Browaeys2020,Ebadi2021} etc., the interaction Hamiltonians to generate unprecedented levels of multipartite entanglement with hundreds of particles generally belong to this family of Eq.\,\eqref{Eq:GeneralH}, which covers a variety of interacting systems of a large number of particles. 

%

%
The key is to find the speed limit of the QFI growth for a quantum metrological resource state $\varrho(t)$ prepared by a unitary evolution $\mathcal{U}(t)$ with the Hamiltonian in Eq.\,\eqref{Eq:GeneralH}, see Fig.\,\ref{Fig:model}a. Starting from a full product state $\sigma=\otimes_{i=1}^N\sigma_i$, the QFI of $\varrho(t)=\mathcal{U}(t)\sigma \mathcal{U}^\dagger(t)$ [cf. Eq.\,\eqref{Eq:QFIdef}] associated with $\mathcal{K}$ can be formulated as \cite{supplement}
\begin{equation}
\label{Eq:QFIReformulation}
F_Q(t)=2c_{\mathrm{WY}}\sum_{i,j}\mathrm{tr}\left([K_i(t),\sqrt{\sigma}]^\dagger [K_j(t),\sqrt{\sigma}]\right), 
\end{equation}
with $c_{\mathrm{WY}}\in[1,2]$ relevant to the Wigner-Yanase skew information \cite{Luo2004}, and $K_i(t)=\mathcal{U}^\dagger(t)K_i \mathcal{U}(t)$. This result reveals an explicit connection between the evolution of QFI  and the spreading as well as interference of on-site operators, see Fig.\,\ref{Fig:model}b. Such a connection enables us to analyze the {\it“growth”} behavior of the QFI  with respect to the preparation time.

In non-relativistic quantum mechanics, by analogy to Einstein's relativity, Lieb-Robinson bound imposes one of the most fundamental restrictions to quantum dynamics, leading to the formation of an {\it “effective light cone”}, $R_{\mathrm{L}}(t)$, that bounds the propagation of quantum information to a finite velocity (i.e. causality in a quantum many-body lattice) \cite{Lieb1972}. Mathematically, the time-dependent commutator between two operators $K_i(t)$ and $K_j$ (i.e. $[K_i(t),K_j]$) decrease rapidly with the graph-theoretic distance $d_{ij}$ separating site $i$ and $j$ outside the light cone, which ensures an approximation of $K_i(t)$ as $K_i(t)\approx K_i(t,i[R])$ \cite{supplement}. Here, $R\gtrsim R_{\mathrm{L}}(t)$ and $K_i(t,i[R])$ represents a projection of $K_i(t)$ onto the subset $i[R]\subset \Lambda$ that is a ball region centered at the site $i$ with radius $R$, see Fig.\,\ref{Fig:model}b. The symbol $A\lesssim B$ represents $A\leq c B$ for some constant $c$. Exploiting such a causality relation, we establish an upper bound for the growth of QFI in many-body quantum lattices as  
\begin{equation}
\label{Eq:QFI_GB}
F_Q(t)\lesssim \kappa c_{\mathrm{WY}}\left[1+ \gamma 2^D  R^D_{\mathrm{L}}(t)\right]N,
\end{equation}
where $\kappa\sim \mathcal{O}(1)$ represents the maximal spectrum width of the local interrogation operators $\{K_i\}$, and $\gamma$ is a positive constant determined by the geometry of a quantum lattice \cite{supplement}. Therefore, given a state preparation time $t$, the maximal QFI that can be achieved is strictly constrained by the geometry and the light cone for a quantum many-body lattice. Such a fundamental constraint determines whether the metrological system can surpass the SQL when accounting for the time overhead of quantum resource state preparation.
{\it Strong precision limit of quantum metrology.---} In quantum metrology, in order to estimate a parameter $\lambda$ encoded in the phase $\vartheta=\lambda \tau$ with $\tau$ the interrogation time of a single measurement, the minimal uncertainty of multiple identical measurements is set via the well-known quantum Cram\'er-Rao bound \cite{supplement},
\begin{equation}
\label{Eq:QCRB}
\delta \lambda\geq  N^{-\Delta/2}  \delta \lambda_{\mathrm{SQL}}\equiv N^{-(1+\Delta)/2}  \frac{1}{\sqrt{T\tau}} 
\end{equation}
Here, $\delta\lambda_{\mathrm{SQL}}\equiv 1/\sqrt{NT\tau}$ represents the standard quantum limit with $T$ the total time of multiple measurements. The exponent $\Delta$ characterizes the quantum metrological enhancement to beat the SQL; specifically, $\Delta=0$ and $1$ correspond to the SQL and Heisenberg scaling respectively. For a sufficiently large quantum many-body system, the preparation time (below denoted as $t_{\mathrm{p}}$) of entangled metrological resource state would generally be much longer than the signal interrogation time (which is limited by the system's coherence time), namely $t_{\mathrm{p}}\gg \tau$, and then the number of measurement repetitions is upper bounded by $T/t_{\mathrm{p}}$. Thus, the enhancing exponent $\Delta$ is dominantly determined by \cite{supplement},
\begin{equation}
\Delta=\log_N \left(\frac{F_Q}{t_{\mathrm{p}}}\cdot\frac{\tau}{N}\right).
\label{eq:Delta}
\end{equation}
Based on the QFI growth bound in Eq.\,\eqref{Eq:QFI_GB}, we are able to determine the maximal exponent $\Delta$ through the scaling behavior of the ratio $(F_Q/t_{\mathrm{p}})$ (hereafter $\tau$ is assumed as unit), and then strengthen the ultimate precision limit of quantum metrology in generic quantum many-body lattices. 
%


%
For short-range interacting systems, namely, the decay of interaction strength is faster than an exponential decay relative to the distance between separated sites, Lieb and Robinson proved in 1972 an effective linear light cone $R_\mathrm{L}(t)\simeq v_{\mathrm{LR}} t$ with $v_{\mathrm{LR}}\sim \mathcal{O}(1)$ the so-called LR velocity \cite{Lieb1972}. By using our result in Eq.\,\eqref{Eq:QFI_GB} that $F_Q\lesssim N R_{\mathrm{L}}^D(t_{\mathrm{p}})\sim N t_{\mathrm{p}}^D$, and thus $t_{\mathrm{p}}\gtrsim (F_Q/N)^{1/D}$,  the metrological enhancing exponent would be upper bounded by \cite{supplement}
\begin{equation}
\label{Eq:Upsilon-1}
\Delta \leq \log_N \left(\frac{F_Q}{N}\right)^{1-\frac{1}{D}}\leq 1-\frac{1}{D},
\end{equation}
where the second inequality stems from the conventional precision limit, i.e. $F_Q\lesssim N^2$  \cite{Boixo2007}.
Particularly, for one dimensional (1D) short-range interacting quantum systems, the result implies that one can only achieve a zero enhancing exponent, i.e. $\Delta=0$, and thereby a surprising precision scaling as $\delta\lambda_{\mathrm{SQL}}$. In this scenario, even if the system might be able to be prepared in a global multipartite entangled state with $F_Q\sim N^2$, the Heisenberg scaling would not be feasible due to the required state preparation time that scales at least linearly with the system size, namely $t_{\mathrm{p}}\gtrsim F_Q/N \sim N$.
\begin{figure}[t]
\centering
\includegraphics[width=88mm]{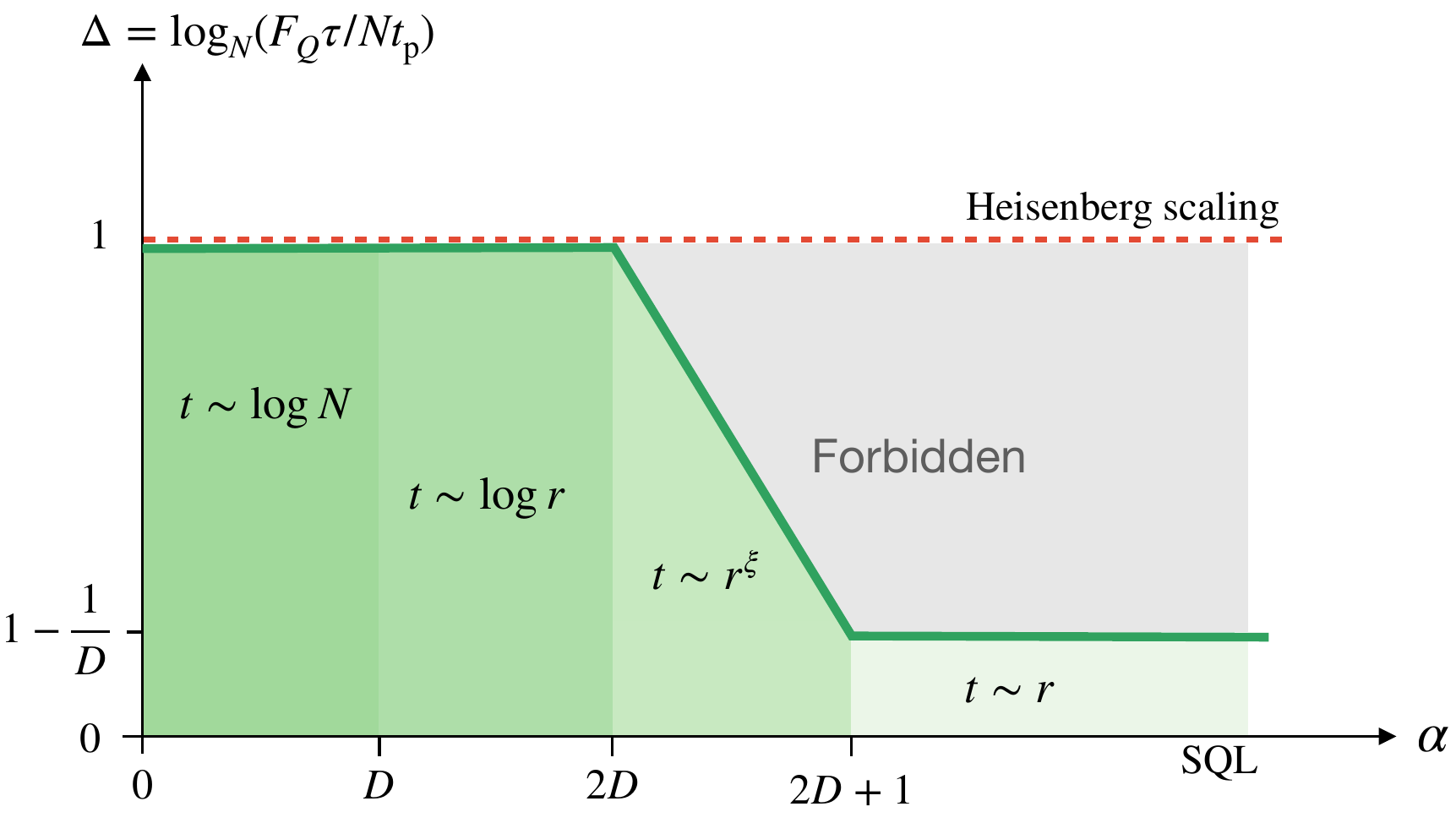}
\caption{{\bf Bound of quantum metrological enhancement.} In the regime of $\alpha>2D+1$, a quantum many-body lattice is governed by a linear light cone ($t\sim r$), resulting in that $\Delta\leq 1-1/D$; while in the regime of $\alpha \leq 2D+1$, “supersonic” information propagation is possible (e.g. polynomial $t\sim r^\xi$ and logarithmic $t\sim \log r$ light cones; or even no light cone, $t\sim \log N$ lower bounds the shortest signaling time between two local sites \cite{Guo2020}), which gives rise to that $\Delta \leq 3-\alpha/D$ for $\alpha \in (2D,2D+1]$ and $\Delta$  approaches the Heisenberg scaling in the limit of large system size for $\alpha \in [0, 2D]$.}
\label{Fig:PL}
\end{figure}
For long-range interacting systems, we consider two-body interactions 
of the form $\left\|h_{i, j}\right\| \lesssim d_{i, j}^{-\alpha}$, where $\{h_{ij}\}$ are bipartite interaction operators, $d_{i,j}$ is the distance from the site $i$ to $j$, $\alpha$ is the power-law decaying exponent. A strictly linear light cone exists for sufficiently large $\alpha$ (i.e. $\alpha>2D+1$) in generic long-range interacting lattices \cite{Kuwahara2020}, and thus the metrological enhancing exponent is also bounded by Eq.\,\eqref{Eq:Upsilon-1}. Nevertheless, as the decaying exponent $\alpha$ becomes smaller, a linear light cone can be broken \cite{Tran2021PRX}, namely $R_\mathrm{L}(t)$ increases nonlinearly as $t$. In the regime of $\alpha\in (2D,2D+1]$, the shape of the effective light cone becomes polynomial and is given by $R_{\mathrm{L}}(t)\simeq c t^{1/\xi}$, 
where $c$ is a constant of $\mathcal{O}(1)$ and $\xi=\alpha-2D$ up to an arbitrarily small constant \cite{Tran2021}. Therefore, our result in Eq.\,\eqref{Eq:QFI_GB}, i.e. $t_{\mathrm{p}}\gtrsim (F_Q/N)^{\xi/D}$, leads to the following metrological enhancing exponent 
\begin{equation}
\label{Eq:Upsilon-2}
\Delta \leq \log_N \left(\frac{F_Q}{N}\right)^{1-\frac{\xi}{D}} \leq 1-\frac{\xi}{D}.
\end{equation}
As $\alpha\to 2D$ and $\xi\to 0^+$, the measurement precision would approach the Heisenberg scaling (i.e. $\Delta=1$). While in the regime of $\alpha \in (D, 2D]$, the LR bounds give rise to a “logarithmic light cone” (i.e. $t\sim \log R$), or equivalently an exponentially fast quantum information propagation \cite{Hastings2006,Tran2021}. Similarly, according to Eq.\,\eqref{Eq:QFI_GB}, this leads to the metrological enhancement as
\begin{equation}
\Delta\leq 1-\log_N\mathrm{polylog}(N).
\label{eq:slimit}
\end{equation}
Note that $\log_N\mathrm{polylog}(N)\to 0$ for $N\to \infty$, therefore the Heisenberg limit becomes asymptotically achievable.  For {\it ultra} long-range interacting systems with $\alpha \in [0,D]$, the notation of quasilocality is broken \cite{Eisert2013,Hauke2013,Richerme2014}; in other words, the causal region defined by an effective light cone might disappear. Nevertheless, the recently generalized LR bound in this scenario \cite{Guo2020} still allows to establish an effective constraint for the metrological enhancing exponent according to Eq.\,\eqref{Eq:QFI_GB}, which implies that the Heisenberg limit is achievable for large $N$ \cite{supplement,Munoz2022}. In Fig.\,\ref{Fig:PL}, we summarize the scaling behavior of these strong precision limits of quantum metrology with respect to the system size for all possible values of $\alpha \geq 0$. In particular, we remark that the interaction regime $\alpha=0$ is favorable to prepare many-body metrological resource states, especially in experimental platforms of Bose-Einstein condensates \cite{Gross2010,Riedel2010,Muessel2014} and cavity-QED systems \cite{Hosten2016,Colombo2022,Greve2022,Leroux2010}.

\begin{figure}[t]
\centering
\includegraphics[width=88mm]{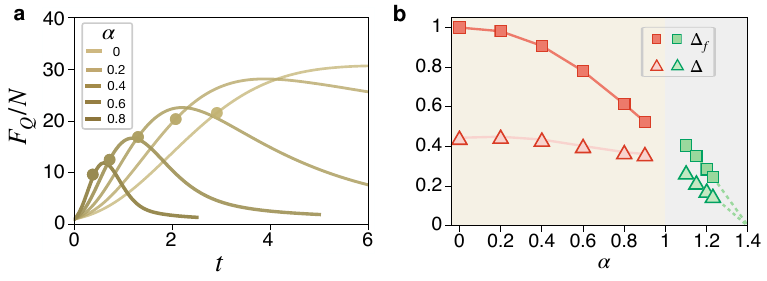}
\caption{{\bf Metrological enhancement by Ising chain with long-range interaction.} {\bf a}, QFI growth with respect to the evolution time for different values of $\alpha$ with $N=60$ sites. Starting from the coherent spin state along the $x$ direction, we maximize the QFI at all times $t$ by optimizing local interrogation operators, $K_i=S_{\hat{n}}$, with $S_{\hat{n}}$ the spin-1/2 operator along the direction $\hat{n}$. The circles mark the optimum of $(F_Q/t_{\mathrm{p}})$ for $t_{\mathrm{p}}>0$. 
{\bf b}, Metrological enhancing exponent $\{\Delta_f = \log_N (F_Q/N),\Delta= \log_N (F_Q\tau/Nt_{\mathrm{p}})\}$ for different values of $\alpha$. We obtain each value of $\Delta$ and $\Delta_f$ by numerically fitting the quantities $(F_Q/t_{\mathrm{p}})$ and $F_Q$ with respect to the system size $N$ \cite{supplement}. 
}
\label{Fig:Ising}
\end{figure}

{\it Design principles for quantum metrological systems.---} Our result suggests that long-range interaction is favorable for achieving the quantum advantage of quantum metrology over the SQL. Below we present an illustrative example of the Ising chain ($D=1$) with the Hamiltonian given by
\begin{equation}
\label{Eq:IsingH}
H_{\mathrm{Ising}}=\frac{1}{\mathcal{N}_\alpha}\sum_{i<j}\frac{4}{|i-j|^\alpha}S_i^z S_j^z,
\end{equation}
Here, $\bm{S}_i$ represents the spin-1/2 operator associated with the site $i$, and $\mathcal{N}_\alpha$ is a normalization factor chosen as $N^{1-\alpha}$ for $\alpha\in[0,1)$, and $1$ for $\alpha>1$ respectively, which stems from the condition of bounded one-site energy in Eq.\,\eqref{Eq:GeneralH} to ensure system energy an extensive quantity \cite{Guo2020}. Despite the fact that all terms of the Ising Hamiltonian commute, rich genuine quantum features (e.g. spin squeezing \cite{Kitagawa1993}, quantum magnetism \cite{Britton2012} etc.) can occur. In Fig.\,\ref{Fig:Ising}a, we investigate the QFI growth dynamics for different values of $\alpha$, and find the maximal point where an optimal enhancing factor $(F_Q/t_{\mathrm{p}})$ [cf. Eq.\eqref{eq:Delta}] is obtained. We note that the all-to-all interaction of $\alpha=0$ shows the slowest QFI growth due to the largest interaction rescaling factor $\mathcal{N}_\alpha$ in Eq.\,\eqref{Eq:IsingH}. When taking the state preparation time into account, the optimal metrological points are no more those maximizing the QFI. As can be seen from Fig.\,\ref{Fig:Ising}b, the metrological enhancing exponents $\Delta$ are smaller than the ones (denoted as $\Delta_f$) without considering the required time for resource state preparation (i.e. solely determined by $F_Q/N$), namely $\Delta<\Delta_f\equiv \log_N(F_Q/N)$. Nevertheless, it still results in an apparent quantum advantage over the SQL (i.e. $\Delta \geq 0.4$) if $\alpha < 1$. In contrast, the SQL can not be surpassed for $\alpha \geq 1.4$, since the maximal QFI would only scale approximately as $F_Q\sim N$ \cite{supplement}.
%

%
Our result also hints that the quantum advantage of quantum metrology can be achieved by exploiting higher-dimensional systems. For a larger dimension $D$, local information can propagate along more directions to spread across the entire lattice, resulting in the speed-up of QFI growth which is evidenced by the power exponent $D$ of $R_{\mathrm{L}}(t)$ [cf. Eq.\,\eqref{Eq:QFI_GB}]. Given a power-law decaying exponent $\alpha$, a larger $D$ would change the shape of the effective light cone, and result in an improved bound for the strong precision limit (see Fig.\ref{Fig:PL}). An important example is the multipartite entanglement generation in quantum systems with $U(1)$ symmetric dipolar interactions, i.e. $H_{X X}=-\sum_{i<j} \left(S_i^x S_j^x+S_i^y S_j^y\right)/d_{i j}^3$. This type of dipolar system can be prominently engineered in dipolar molecules \cite{Moses2017}, solid-state spin ensembles \cite{Choi2017,Zheng2022}, Rydberg-atom arrays \cite{Browaeys2020}, etc. In the 1D case, the linear light cone (i.e. $\alpha=3 \geq 2D+1$ with $D=1$) prevents the advantage of quantum metrology (namely $\Delta = 0$) due to the time overhead for state preparation. When we extend to the higher-dimensional system (e.g. $D=2$), $\alpha$ would fall into the regime of $(D,2D)$, and the linear light cone fails to hold, which would make it possible to surpass the SQL according to our result, see e.g. \cite{Comparin2022}. By further engineering the forms of long-range interactions (e.g. by designing quantum gate sequences etc.) of a given power-law decaying exponent $\alpha$, it is possible to achieve the asymptotic Heisenberg limit \cite{Tran2021PRX,Tran2021}, and thus confirms the strong precision limit in Eq.\,\eqref{eq:slimit}.

{\it Conclusion $\&$ prospects.---} To summarize, we reveal a connection between Lieb-Robinson light cone and entanglement-enhanced quantum metrology, and present a fundamental constraint for the QFI growth in generic quantum many-body lattices with bounded one-site energy. This bound allows us to establish strong precision limits for quantum metrology that include the exceptional time overhead of quantum resource state preparation. We remark that such overhead in entanglement-enhanced quantum metrology stands out as a universal and physically fundamental constraint imposed by quantum mechanical laws, as compared with experimentally technical ones which may be overcome by the advance of technologies. Our result allows us to identify the crucial features of quantum many-body systems for entanglement-enhanced quantum metrology, and thus provides guiding principles for the design of many-body quantum metrological systems. A remaining complementary aspect would be taking the readout process into account as well. We also note that it might be possible in realistic experimental scenarios for the existence of other specific time overheads, and the theoretically interesting question would be identifying the universal part to further tighten the present precision limit.

Given the increasingly significant roles of QFI in many-body physics \cite{Frerot2018,Rams2018,Niezgoda2021,Smith2016,Tomasi2019,Garttner2018,Desaules2022}, the established general bound of QFI growth [cf. Eq.\,\eqref{Eq:QFI_GB}] might also provide interesting insights into relevant quantum dynamics \cite{supplement}. For example, our result predicts that, by starting from a full product state and engineering short-range interactions in a 1D lattice, the minimal time to prepare a metrologically useful, $k$-partite entangled state (i.e. $F_Q/N\geq k$ \cite{Hauke2016}) would be lower bounded by $t \gtrsim k/v_{\mathrm{LR}}$. The present framework can be extended to dynamical quantum metrological protocols as well, where the parameter is encoded via a dynamical evolution \cite{Hayes2018,Chu2020,Chu2021} governed by a Hamiltonian of the form, $\mathcal{H}=H+\lambda \mathcal{K}$, with $H$ given by Eq.\,\eqref{Eq:GeneralH}. For 1D short-range interacting systems, we find that the interrogation time shall satisfy $\tau\gtrsim N$ to achieve the Heisenberg limit  \cite{supplement}, which further sets conditions for the system's coherence time.

{\it Acknowledgements.--- } We thank Lijian Zhang for reading our manuscript and providing helpful comments. We also thank Zhendong Zhang and Andrew Guo for  valuable discussions about Bose-Einstein condensates and Lieb-Robinson bounds respectively. The work is supported by the National Natural Science Foundation of China (12161141011), the National Key R$\&$D Program of China (2018YFA0306600), the Interdisciplinary Program of the Wuhan National High Magnetic Field Center (WHMFC202106), and Shanghai Key Laboratory of Magnetic Resonance (East China Normal University). Y.-M.C. is also supported by a China Postdoctoral Science Foundation Grant (No. 2022M721256).
%

%

\bibliography{reference}
 

\section*{Supplementary material}

# Supplementary Materials

**CONTENTS**

## 1. Connection between QFI growth and local operator spreading

By noticing that

$$p_n + p_m \leq (\sqrt{p_n} + \sqrt{p_m})^2 \leq 2(p_n + p_m), \tag{S1}$$

for $p_n > 0$ and $p_m > 0$, one can find that

$$(\sqrt{p_n} - \sqrt{p_m})^2 \leq \frac{(p_n - p_m)^2}{p_n + p_m} \leq 2(\sqrt{p_n} - \sqrt{p_m})^2. \tag{S2}$$

Based on the definition of QFI in Eq. (1) in the main text, we have

$$2\sum_{n,m}(\sqrt{p_n} - \sqrt{p_m})^2 |\langle n|\mathcal{K}|m\rangle|^2 \leq F_Q[\varrho, \mathcal{K}] \leq 4\sum_{n,m}(\sqrt{p_n} - \sqrt{p_m})^2 |\langle n|\mathcal{K}|m\rangle|^2. \tag{S3}$$

By further introducing a coefficient $c_{\mathrm{WY}} \in [1, 2]$, the following expression for the QFI can be obtained,

$$F_Q[\varrho, \mathcal{K}] = 2c_{\mathrm{WY}} \sum_{n,m}(\sqrt{p_n} - \sqrt{p_m})^2 |\langle n|\mathcal{K}|m\rangle|^2. \tag{S4}$$

Therefore, the QFI of the prepared metrological state at an evolution time $t$, i.e. $\varrho(t) = \mathcal{U}(t)\sigma\mathcal{U}^\dagger(t)$ with $\sigma$ an initial product state, can be formulated as

$$\begin{aligned} F_Q[\varrho(t), \mathcal{K}] &= 2c_{\mathrm{WY}} \sum_{n,m}(\sqrt{p_n} - \sqrt{p_m})^2 |\langle n|\mathcal{K}|m\rangle|^2 \\ &= 2c_{\mathrm{WY}} \sum_{n,m} \langle n|[\mathcal{K}, \sqrt{\varrho(t)}]^\dagger|m\rangle \langle m|[\mathcal{K}, \sqrt{\varrho(t)}]|n\rangle \\ &= 2c_{\mathrm{WY}} \operatorname{tr}\left([\mathcal{K}, \sqrt{\varrho(t)}]^\dagger [\mathcal{K}, \sqrt{\varrho(t)}]\right) \\ &= 2c_{\mathrm{WY}} \operatorname{tr}\left([\mathcal{K}(t), \sqrt{\sigma}]^\dagger [\mathcal{K}(t), \sqrt{\sigma}]\right) \\ &= 2c_{\mathrm{WY}} \sum_{i,j} \operatorname{tr}\left([K_i(t), \sqrt{\sigma}]^\dagger [K_j(t), \sqrt{\sigma}]\right), \end{aligned} \tag{S5}$$

Here, the interrogation operator is $\mathcal{K}(t) = \mathcal{U}^\dagger(t)\mathcal{K}\mathcal{U}(t) = \sum_i K_i(t)$, where $K_i(t) \equiv \mathcal{U}^\dagger(t)K_i\mathcal{U}(t)$ represents the time evolution of local interrogation operators in the Heisenberg picture. The last equality [namely Eq. (3) in the main

text] implies that the QFI dynamics to prepare a metrological resource state is deeply related to the spreading of local operators. We define the component of the QFI collaboratively contributed by the site $i$ and $j$ as

$$\alpha_{ij}(t) \equiv \mathrm{tr}\left([K_i(t), \sqrt{\sigma}]^\dagger [K_j(t), \sqrt{\sigma}]\right), \tag{S6}$$

and the QFI can then be written as

$$F_Q[\varrho(t), \mathcal{K}] = 2c_{\mathrm{WY}} \sum_{i,j} \alpha_{ij}(t). \tag{S7}$$

Generally, by utilizing Cauchy–Schwarz inequality, we obtain that

$$\begin{aligned}|\alpha_{ij}(t)|^2 &\leq \mathrm{tr}\left([K_i(t), \sqrt{\sigma}]^\dagger [K_i(t), \sqrt{\sigma}]\right) \mathrm{tr}\left([K_j(t), \sqrt{\sigma}]^\dagger [K_j(t), \sqrt{\sigma}]\right) \\ &\leq \frac{1}{4} F_Q[\varrho, K_i] F_Q[\varrho, K_j] \\ &\leq \frac{1}{4} \|K_i\|_{\mathrm{s}}^2 \|K_j\|_{\mathrm{s}}^2. \end{aligned} \tag{S8}$$

The second inequality results from a similar relation as Eq. (S5) between the QFI and the local operator $K_i$, namely

$$F_Q[\varrho(t), K_i] = 2c_{\mathrm{WY}} \mathrm{tr}\left([K_i(t), \sqrt{\sigma}]^\dagger [K_i(t), \sqrt{\sigma}]\right) = 2c_{\mathrm{WY}} \alpha_{ii}(t), \tag{S9}$$

which directly yields that

$$\alpha_{ii}(t) \leq \frac{1}{2} F_Q[\varrho(t), K_i]. \tag{S10}$$

While in the third inequality, we use the fact that $F_Q[\varrho, K_i] \leq \|K_i\|_{\mathrm{s}}^2$ (cf. Ref. [S1]), with the semi-norm $\|\bullet\|_{\mathrm{s}}$ taking the spectrum width, namely the difference between the largest and smallest eigenvalues. By further defining $\kappa \equiv \max_{i \in \Lambda} \|K_i\|_{\mathrm{s}}^2$ as the maximum spectrum width of the local interrogation operators $\{K_i\}$, we conclude that $|\alpha_{ij}(t)| \leq \kappa/2 \sim \mathcal{O}(1)$ for arbitrary $i$ and $j$.

3

## 2. Causality relation for local operator spreading

In this section, we sketch the proof for the projection of the time-evolved local operator [subject to the Hamiltonian in Eq. (2) of the main text], $K_i(t) = \mathcal{U}^\dagger(t) K_i \mathcal{U}(t)$. Mathematically, we define the projection operation onto its neighboring ball region denoted by $i[R]$ as

$$K_i(t, i[R]) \equiv \frac{1}{\text{tr}_{i[R]^c} \hat{\mathbb{1}}_{i[R]^c}} \text{tr}_{i[R]^c}[K_i(t)] \otimes \hat{\mathbb{1}}_{i[R]^c}, \tag{S11}$$

where $i[R]^c$ denotes the complementary set of $i[R]$, i.e. $i[R]^c = \Lambda \setminus i[R]$. Following Ref. [S2], let $U$ be a unitary operator acting on $i[R]^c$ and $\mu(U)$ be the Haar measure for $U$, one can rewrite $K_i(t, i[R])$ in the following form

$$K_i(t, i[R]) = \int d\mu(U) U K_i(t) U^\dagger, \tag{S12}$$

and therefore

$$\|K_i(t) - K_i(t, i[R])\| \leq \int d\mu(U) \|[K_i(t), U]\| \leq \max_U \|[K_i(t), U]\|, \tag{S13}$$

where $\max_U$ accepts all unitary operators $U$ supported on $i[R]^c$. By further using the standard commutator form of Lieb-Robinson bound for generic quantum many-body lattices, one can find that

$$\|K_i(t) - K_i(t, i[R])\| \leq \max_U \|[K_i(t), U]\| \leq C(t, R). \tag{S14}$$

Here, $C(t, R)$ is a function that takes the forms of: (i) $C(t, R) \sim e^{vt-R}$ and $v \sim \mathcal{O}(1)$ for short-range interaction [S3]; (ii) $C(t, R) \sim t^a/R^b$ with $a = (\alpha - D)/(\alpha - 2D)$ and $b = \alpha - 2D$ for long-range interactions with $\alpha > 2D$ [S4]; the effective causal region can be obtained by requiring that $C(t, R) \leq \mathcal{O}(1)$, which results in a light cone of the form $R_L(t) \sim t^{a/b}$ and thus $\xi = b/a = \alpha - 2D$ in the main text; (iii) $C(t, R) \sim e^{vt}/R^\alpha$ and $v \sim \mathcal{O}(1)$ for long-range interactions

4

with $\alpha \in (D, 2D]$ [S5]. The right-hand side of Eq. (S14) would decay to zero by requiring that $R \gtrsim R_{\mathrm{L}}(t)$ (up to a constant factor), where $R_{\mathrm{L}}(t)$ represents the effective light cone. In this sense [S6], one obtains that

$$K_i(t) \simeq K_i(t, i[R]) \quad \text{if} \quad R \gtrsim R_{\mathrm{L}}(t), \tag{S15}$$

and below we further abbreviate it as $K_{i[R]}(t) \equiv K_i(t, i[R])$ for simplicity.

### 3. Proof of the bound for QFI growth

Based on the causality relation in the above section, the component of the QFI collaboratively contributed by the site $i$ and $j$ can be approximately rewritten as

$$\alpha_{ij}(t) \simeq \mathrm{tr}\left([K_{i[R]}(t), \sqrt{\sigma}]^\dagger [K_{j[R]}(t), \sqrt{\sigma}]\right) \quad \text{if} \quad R \gtrsim R_{\mathrm{L}}(t). \tag{S16}$$

Importantly, if the distance between the site $i$ and $j$ is outside the light cone, i.e. $d_{ij} \geq 2R \gtrsim 2R_L(t)$, the two ball regions $K_{i[R]}(t)$ and $K_{j[R]}(t)$ would fail to intersect with each other (see Fig. 1b in the main text), and one can derive that

$$\begin{aligned}
\alpha_{ij}(t) &= \mathrm{tr}\left(\sqrt{\sigma_{\Lambda\setminus i[R]}}[K_{i[R]}(t), \sqrt{\sigma_{i[R]}}]^\dagger [K_{j[R]}(t), \sqrt{\sigma_{j[R]}}]\sqrt{\sigma_{\Lambda\setminus j[R]}}\right) \\
&= \mathrm{tr}\left(\sqrt{\sigma_{i[R]}}[K_{i[R]}(t), \sqrt{\sigma_{i[R]}}]^\dagger [K_{j[R]}(t), \sqrt{\sigma_{j[R]}}]\sqrt{\sigma_{j[R]}}\sigma_{\Lambda\setminus(i[R]\cup j[R])}\right) \\
&= \mathrm{tr}\left(\sqrt{\sigma_{i[R]}}[K_{i[R]}(t), \sqrt{\sigma_{i[R]}}]^\dagger\right) \mathrm{tr}\left([K_{j[R]}(t), \sqrt{\sigma_{j[R]}}]\sqrt{\sigma_{j[R]}}\right) \\
&= 0,
\end{aligned} \tag{S17}$$

where $\sigma_A$ denotes the projection of $\sigma$ onto the region $A \subset \Lambda$. The second equality results from that

$$\sqrt{\sigma_{\Lambda\setminus i[R]}}\sqrt{\sigma_{\Lambda\setminus j[R]}} = \sqrt{\sigma_{i[R]}}\sqrt{\sigma_{j[R]}}\sigma_{\Lambda\setminus(i[R]\cup j[R])}, \tag{S18}$$

5

due to the non-intersection of the two sets $i[R]$ and $j[R]$. The fourth equality in Eq. (S17) can be obtained by noticing the identity that $\text{tr}([A, B]B) = 0$ with $A$ and $B$ arbitrary Hermitian operators. Consequently, the QFI (Eq.S7) can be bounded as

$$F_Q = 2c_{\text{WY}} \sum_i \left[ \alpha_{ii}(t) + \beta_i(t) \sum_{j \in i[2R] \setminus i} 1 \right] \quad \text{(S19)}$$
$$\leq 2c_{\text{WY}} \sum_i \left[ \alpha_{ii}(t) + \beta_i(t) \gamma 2^D R^D \right]$$

with $\beta_i(t) \equiv \sum_{j \in i[2R] \setminus i} \text{Re}[\alpha_{ij}(t)] / \sum_{j \in i[2R] \setminus i} 1 \leq \kappa/2$. In the second inequality of Eq. (S19), we have used a geometry assumption that $\max_{i \in \Lambda} |i[R]| \leq \gamma R^D$, where the introduced geometric parameter $\gamma$ is determined based on the lattice structure alone. By further defining that $\alpha_t \equiv \max_{i \in \Lambda} \alpha_{ii}(t)$ as well as $\beta_t \equiv \max_{i \in \Lambda} \beta_i(t)$, and taking $R = R_\text{L}(t)$, we arrive at

$$F_Q \lesssim 2c_{\text{WY}}[\alpha_t + \beta_t 2^D \gamma R_\text{L}^D(t)]N. \quad \text{(S20)}$$

This result straightforwardly gives rise to Eq. (4) in the main text using $\{|\alpha_t|, |\beta_t|\} \leq \kappa/2$.

### 4. Proof of the strong precision limit of quantum metrology

The celebrated quantum Cramér-Rao bound [S7] states that the ultimate measurement precision via local unbiased estimates is lower bounded by

$$\delta \lambda \geq \frac{1}{\sqrt{\nu F_Q \tau^2}}, \quad \text{(S21)}$$

where $\nu$ is the number of identical measurement repetitions and is constrained by the total measurement time $T$, i.e. $\nu \leq T/(\tau + t_\text{p})$. Here, $t_\text{p}$ and $\tau$ represent

6

the required time for single-run metrological state preparation and signal interrogation respectively. As a result, the above precision limit can be reformulated as

$$\delta\lambda \geq \sqrt{\frac{N(1+t_{\rm p}/\tau)}{F_Q}}\delta\lambda_{\rm SQL} \equiv N^{-\frac{\Delta}{2}}\delta\lambda_{\rm SQL}, \qquad (S22)$$

where $\delta\lambda_{\rm SQL} \equiv 1/\sqrt{NT\tau}$ represents the standard quantum limit (SQL). By making the natural assumption that $t_{\rm p} \gg \tau$ for large many-body quantum systems, one would obtain Eq. (5) with the metrological enhancing exponent [Eq. (6) in the main text] defined as

$$\Delta = -2\log_N\left(\sqrt{\frac{N(1+t_{\rm p}/\tau)}{F_Q}}\right) \approx \log_N\left(\frac{F_Q}{t_{\rm p}}\cdot\frac{\tau}{N}\right). \qquad (S23)$$

Without loss of generality, we can set $\tau = 1$ as a unit and focus on the scaling behavior of $F_Q/t_{\rm p}$ with respect to $N$.

Next, we exploit the established bound of the QFI growth [cf. Eq. (4) in the main text], i.e. $F_Q/N \lesssim R_{\rm L}^D(t_{\rm p})$ (up to some constant coefficients) to derive the strong precision limits set by Eqs. (7-9) in the main text. For short-range interacting system, i.e. $R_{\rm L}(t_{\rm p}) \sim t_{\rm p}$ [S3, S6], using our result in Eq. (4) of the main text, we can conclude that, in order to achieve the QFI density of $F_Q/N \sim N^{\Delta_f}$ with the assumption that $\Delta_f \in [0,1]$, a metrological state preparation time at least on the order of

$$t_{\rm p} \sim \left(\frac{F_Q}{N}\right)^{\frac{1}{D}} \sim N^{\frac{\Delta_f}{D}}, \qquad (S24)$$

would be required. Hence, according to Eq. (S22), the metrological enhancing exponent in this scenario is given by

$$\Delta = \log_N N^{\left(1-\frac{1}{D}\right)\Delta_f} = \left(1-\frac{1}{D}\right)\Delta_f \leq 1-\frac{1}{D}, \qquad (S25)$$

namely Eq. (7) in the main text. While for long-range interacting systems: (i) if $\alpha \in (2D, 2D+1]$, the light cone is given by $R_L(t_p) \sim t_p^{1/\xi}$ [S4], and then Eq. (4) in the main text allows us to find that the minimal state preparation time to reach $F_Q/N \sim N^{\Delta_f}$ needs to satisfy

$$t_p \sim \left(\frac{F_Q}{N}\right)^{\frac{\xi}{D}} \sim N^{\frac{\xi}{D}\Delta_f}, \tag{S26}$$

which further leads to Eq. (8) in the main text, i.e.

$$\Delta = \log_N\left(\frac{N^{\Delta_f}}{t_p}\right) = \left(1 - \frac{\xi}{D}\right)\Delta_f \leq 1 - \frac{\xi}{D}. \tag{S27}$$

(ii) if $\alpha \in (D, 2D]$, then $R_L(t_p) \sim e^{t_p}$ [S5, S8]. Similarly, the state preparation time shall be larger than

$$t_p \sim \log N^{\Delta_f/D} \sim \text{polylog}(N), \tag{S28}$$

and thus using Eq. (S22) we have

$$\Delta \leq 1 - \log_N \text{polylog}(N), \tag{S29}$$

namely Eq. (9) in the main text.

For *utra* long-range interacting systems with the interaction decaying exponent $\alpha \in [0, D]$, while still under increasing investigations for the generalized Lieb-Robinson bounds, the recently developed one [S9] have established that

$$\|K_i(t) - K_i(t, i[R])\| \leq C(t, R) \lesssim |i[R]^c| \frac{e^{vt} - 1}{N^{1-\alpha/D} R^\alpha}. \tag{S30}$$

Generally, this bound can not define an effective causal region; for example, taking $\alpha = 0$, one obtains that $C(t, R) \lesssim (e^{vt} - 1)$ for $|i[R]^c| \sim N$ (i.e. $0 \leq R \lesssim N^{1/D}$), which fails to define $R_L(t)$. However, one can still get interesting conclusions according to our results. If $t$ is very small such that

8

$C(t, R) \ll 1$ for arbitrary $R$, $K_i(t)$ would still be a local operator (i.e. almost supported on-site $i$) and the QFI $F_Q(t) \sim N$. In order to achieve the QFI of $F_Q \sim N^2$, we require $R \sim N^{1/D}$ according to Eq. (4) in the main text, which further leads to that the state preparation time should satisfy $t_p \geq \mathcal{O}(1)$ by using the above result. Subsequently, according to Eq. (S22), the Heisenberg limit is achievable for large system size ($N \gg 1$). We note that the factor $|i[R]^c|$ in $C(t, R)$ of Eq. (S30) does not exist in previous LR bounds [S2, S5]. Whether such a factor may also be dropped in the present scenario is still an open question under investigation. If this would be the case, the minimal time to achieve $F_Q \sim N^2$ would be lower bounded by $t_p \gtrsim \log(N)$, which further gives rise to

$$\Delta \leq 1 - \log_N \text{polylog}(N), \tag{S31}$$

also approaching the Heisenberg scaling for large $N$. This result is evidenced by the very recent work about optimal state preparation at $\alpha = 0$ using tools from classical nonlinear dynamics [S10].

### 5. Behavior of the QFI growth of Ising model

The pure Ising model for demonstrating the metrological enhancement of long-range interacting systems is exactly solvable. Below, we provide the analytical formulas to calculate the QFI dynamics. Starting from the coherent spin state along the $x$ direction, the maximal QFI at a given evolution time $t$ (by optimizing the interrogation operator $S_{\hat{n}}$ that points along the direction $\hat{n}$) can be obtained as follows

$$F_Q(t) = N + \max_\phi \sum_{i \neq j} \left[ \sin^2(\phi) C_{ij}^y(t) + \frac{1}{2} \sin(2\phi) C_{ij}^{yz}(t) \right], \tag{S32}$$

9

where

$$C_{ij}^y(t) = \frac{1}{2} \prod_{k \neq i,j} \cos\left[2(J_{ik} - J_{jk})t\right] - \frac{1}{2} \prod_{k \neq i,j} \cos\left[2(J_{ik} + J_{jk})t\right], \qquad (S33)$$

and

$$C_{ij}^{yz}(t) = -\sin(2J_{ij}t)\left(\prod_{k \neq i,j} \cos(2J_{ik}t) + \prod_{k \neq i,j} \cos(2J_{jk}t)\right), \qquad (S34)$$

with $J_{ij} \equiv 4/(\mathcal{N}_\alpha |i-j|^\alpha)$. For $\alpha = 0$ and $J_{ij} = 4/N$, the above formulas reduce to the well-known results for spin squeezing under one-axis-twisting quantum dynamics [S11, S12], namely

$$F_Q(t) = N + N(N-1)\left(A + \sqrt{A^2 + B^2}\right)/4, \qquad (S35)$$

where $A = 1 - [\cos(2u)]^{N-2}$, and $B = 4\sin(u)[\cos(u)]^{N-2}$ with $u = 8t/N$.

Based on these analytical formulas of the QFI at any evolution time $t$, the QFI dynamics for a given system size $N$ can be determined numerically. In Fig. S1, we depict the scaling behavior of the maximal QFI by setting the value of the interaction decaying exponent $\alpha$ as 0.5 and 1.4 respectively [cf. Eq. (10) in the main text]. Importantly, our numerical result suggests that the maximal QFI for $\alpha \geq 1.4$ would not surpass the SQL.

### 6. QFI growth and many-body dynamics

The quantum Fisher information plays increasingly significant roles in many-body physics, such as quantum phase transitions [S13, S14], many-body nonlocality [S15], many-body localization [S16, S17], multiple-quantum coherence [S18] and quantum many-body scars [S19], etc. Therefore, the established general bound of QFI growth [cf. Eq. (4) in the main text] might also provide

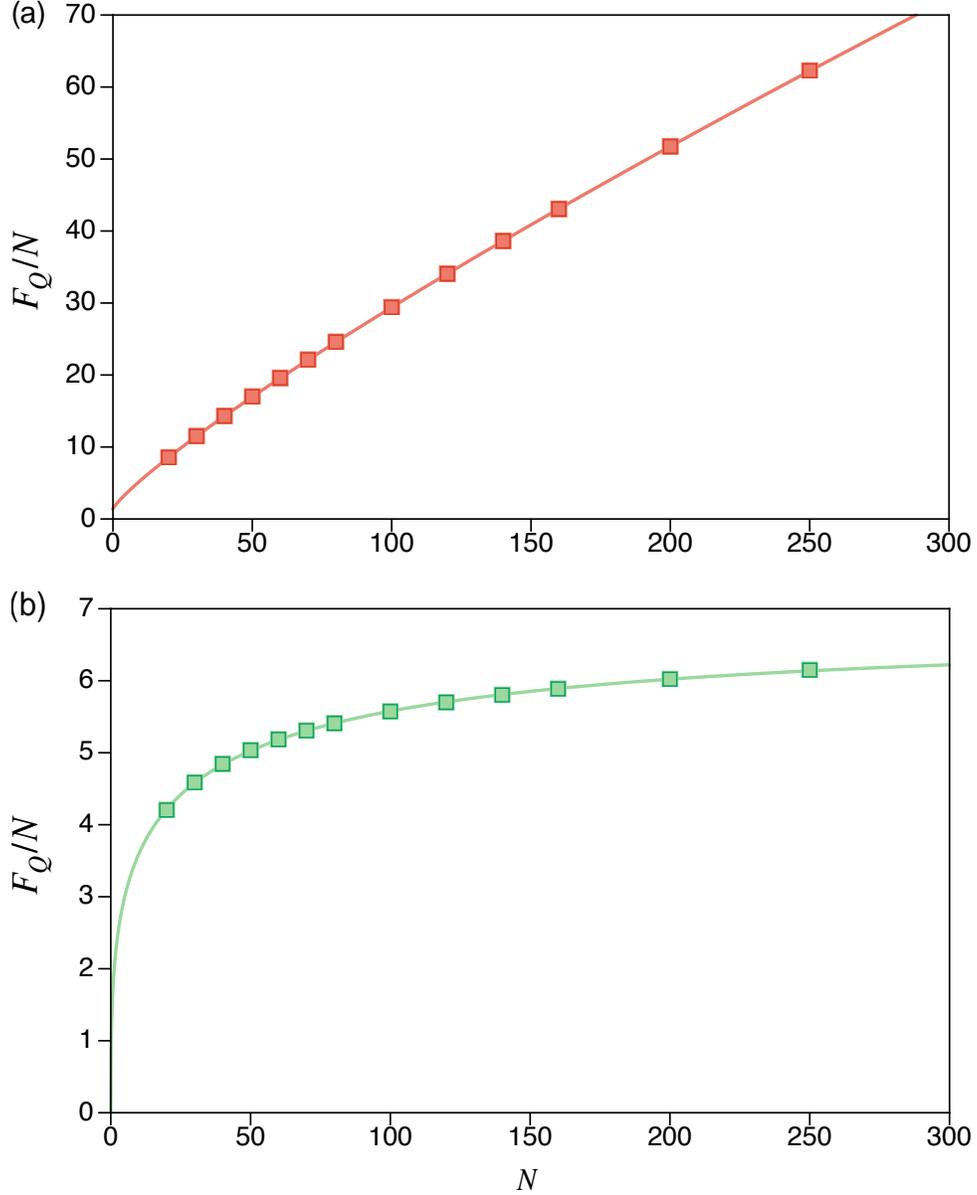

FIG. S1. **Scaling behavior of the QFI for long-range Ising chain.** Starting from the coherent spin state along the $x$ direction, we maximize the QFI at all times $t$ by optimizing local interrogation operators, $K_i = S_{\hat{n}}$, with $S_{\hat{n}}$ the spin-1/2 operator along the direction $\hat{n}$. Each square marks the optimum of $F_Q$ for a specific value of $N$. **a**, Polynomial increase of the QFI at $\alpha = 0.5$ as $F_Q/N = a + bN^c$ with $a = 1.41$, $b = 0.57$ and $c = 0.85$ (red curve). **b**, Saturation of the QFI for $\alpha = 1.4$. The maximal QFI is well fitted by the green curve, i.e. of the form $F_Q/N = a[1 - e^{-(N/b)^c}]$ with $a \approx 6.65$, $b \approx 19.76$ and $c \approx 0.37$. Therefore, for large-size systems, the QFI would saturate as $F_Q \sim N$.

interesting insights into relevant quantum dynamics. For example, considering the GHZ state preparation through adiabatic strategy from the paramagnetic phase of a transverse field Ising chain (namely with an initial state of coherent spin state), we can immediately conclude that the minimal evolution time of at least $t \sim F_Q/v_{\text{LR}}N \sim N$ is required according to the present bound of QFI growth. This is consistent with $\mathcal{O}(N^z)$ given by the adiabatic theorem (here $z$ represents the dynamical critical exponent and $z = 1$ in this case) [S14]. More generally, our result predicts that, by starting from a full product state and engineering short-range interactions in a 1D lattice, the minimal time to prepare a metrologically useful, $k$-partite entangled state (i.e. $F_Q/N \geq k$ [S20]) would be lower bounded by $t \gtrsim k/v_{\text{LR}}$.

### 7. Extended to dynamical quantum metrological protocols

Our framework can be extended to dynamical quantum metrological protocols [S21–S23], where the parameter information is encoded via a dynamical evolution governed by a Hamiltonian of the form,

$$\mathcal{H} = H + \lambda \mathcal{K}, \tag{S36}$$

with $H$ given by Eq. (2) of the main text. In this framework, the corresponding time-evolved local operator $K_i(t)$ in Eq. (3) of the main text would be generalized by the time average

$$\bar{K}_i(\tau) = \int_0^\tau K_i(s) \mathrm{d}s, \tag{S37}$$

over a single run of interrogation and then the QFI growth would be bounded by the effective light cone as

$$F_Q/N \lesssim \int_0^\tau \mathrm{d}t_1 \int_0^\tau \mathrm{d}t_2 [R_{\text{L}}(t_1) + R_{\text{L}}(t_2)]^D. \tag{S38}$$

12

For 1D short-range interacting systems with $R_{\text{L}}(t) \sim t$, this integral would yield that $F_Q \lesssim N\tau^3$, which implies that achieving the Heisenberg limit (i.e. $F_Q \sim N^2\tau^2$ [S1, S14]) via dynamical quantum metrological protocols would require the interrogation time $\tau \gtrsim N$. This result then further sets conditions for the system's coherence time in order to reach the Heisenberg limit.

---



\end{document}